\documentclass{llncs}
\usepackage{graphicx}
\usepackage{amsmath}
\usepackage{amssymb}
\usepackage[ruled, noline, algo2e, noend]{algorithm2e}
\usepackage[T1]{fontenc}
\usepackage{enumerate}
\usepackage[font=small, format=hang, labelfont=bf]{caption}
\usepackage[]{caption}
\usepackage{subfig}
\usepackage{color}
\usepackage{arcs}

\SetArgSty{}  \SetKw{KwOr}{or}
\SetKw{KwAnd}{and} \SetKw{KwInvariant}{invariant:}
\SetKw{KwReturn}{return} \SetKw{KwDownTo}{down to}
\SetKwInOut{Input}{input} \SetKwInOut{Output}{output}
\SetKwInOut{Require}{require} \SetKwComment{Comment}{start}{end}
\SetKwIF{If}{ElseIf}{Else}{if}{then}{else if}{else}{endif}

\newenvironment{sketchofproof}{\noindent\textit{Sketch of
proof.}}{\mbox{}\hfill\qed\par\medskip}
\newcommand{\NP}{\mathcal{NP}}

\newcommand{\eat}[1] {{}}

\newcommand{\GSimRAC}{\mathrm{GSimRAC}}
\newcommand{\GDual}{\mathrm{GDual\text{-}GSimRAC}}

\begin{document}

\title{Geometric Simultaneous RAC Drawings of Graphs}

\author{E. N. Argyriou \inst{1}, M. A.\ Bekos \inst{1}, M. Kaufmann \inst{2} and A. Symvonis \inst{1}}

\authorrunning{E.N. Argyriou, M.A. Bekos, Michael Kaufmann, A. Symvonis}

\tocauthor{Evmorfia N. Argyriou, Michael A. Bekos, Michael Kaufmann,
Antonios Symvonis}

\authorrunning{E.N. Argyriou, M.A. Bekos, M. Kaufmann, A. Symvonis}

\institute{%
    School of Applied Mathematical \& Physical Sciences,\\
    National Technical University of Athens, Greece.\\
    \email{$\{$fargyriou,mikebekos,symvonis$\}$@math.ntua.gr}
    \and
    University of  T\"ubingen, Institute for Informatics, Germany.\\
    \email{mk@informatik.uni-tuebingen.de}
}

\maketitle

\begin{abstract}
In this paper, we introduce and study \emph{geometric simultaneous
RAC drawing problems}, i.e., a combination of problems on geometric
RAC drawings and geometric simultaneous graph drawings. To the best
of our knowledge, this is the first time where such a combination is
attempted.
\end{abstract}

\section{Introduction}
\label{sec:introduction}
A \emph{geometric right-angle crossing drawing} (or \emph{geometric
RAC drawing}, for short) of a graph is a straight-line drawing in
which every pair of crossing edges intersects at right-angle. A
graph which admits a geometric RAC drawing is called
\emph{right-angle crossing graph} (or \emph{RAC graph}, for short).
Motivated by cognitive experiments of Huang et al.\
\cite{Hu07,HHE08}, which indicate that the negative impact of an
edge crossing on the human understanding of a graph drawing is
eliminated in the case where the crossing angle is greater than
seventy degrees, RAC graphs were recently introduced in \cite{DEL09}
as a response to the problem of drawing graphs with optimal crossing
resolution.

\emph{Simultaneous graph drawing} deals with the problem of drawing
two (or more) planar graphs on the same set of vertices on the
plane, such that each graph is drawn planar\footnote{In the graph
drawing literature, the problem is known as ``simultaneous graph
drawing with mapping''. For simplicity, we use the term
``simultaneous graph drawing''.} (i.e., only edges of different
graphs are allowed to cross). The \emph{geometric} version restricts
the problem to straight-line drawings. Besides its independent
theoretical interest, this problem arises in several application
areas, such as software engineering, databases and social networks,
where a visual analysis of evolving graphs, defined on the same set
of vertices, is useful.

Both problems mentioned above are active research topics in the
graph drawing literature and positive and negative results are known
for certain variations (refer to Section~\ref{sec:related-work}). In
this paper, we present the first combinatorial results for the
\emph{geometric simultaneous RAC drawing problem} (or \emph{GSimRAC}
drawing problem, for short), i.e., a combination of both problems.
Formally, the $\GSimRAC$ drawing problem can be stated as follows:
Let $G_1=(V,E_1)$ and $G_2=(V,E_2)$ be two planar graphs that share
a common vertex set but have disjoint edge sets, i.e., $E_1
\subseteq V \times V$, $E_2 \subseteq V \times V$ and $E_1 \cap E_2
= \emptyset$. The main task is to place the vertices on the plane so
that, when the edges are drawn as straight-lines, (i)~each graph is
drawn planar, (ii)~there are no edge overlaps, and, (iii)~crossings
between edges in $E_1$ and $E_2$ occur at right-angles. Let $G=(V,
E_1 \cup E_2)$ be the graph induced by the union of $G_1$ and $G_2$.
Observe that $G$ should be a RAC graph, which implies that $|E_1
\cup E_2| \leq 4|V|-10$ \cite{DEL09}. We refer to this relationship
as the \emph{RAC-size constraint}.

\begin{figure}[t!hb]
  \centering
  \begin{minipage}[b]{.28\textwidth}
     \centering
     \subfloat[\label{fig:nonrac-graph}{}]
     {\includegraphics[width=.7\textwidth]{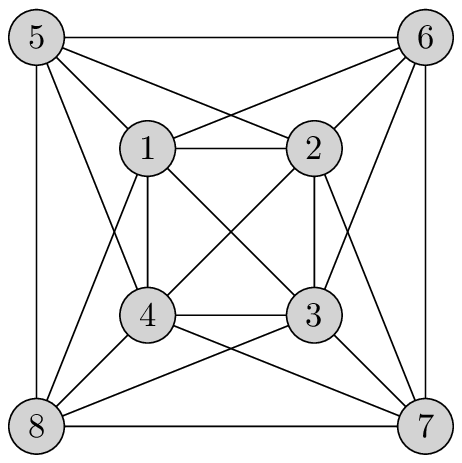}}
  \end{minipage}
  \hfill
  \begin{minipage}[b]{.28\textwidth}
     \centering
     \subfloat[\label{fig:nonrac-graph-decomposition}{}]
     {\includegraphics[width=.7\textwidth]{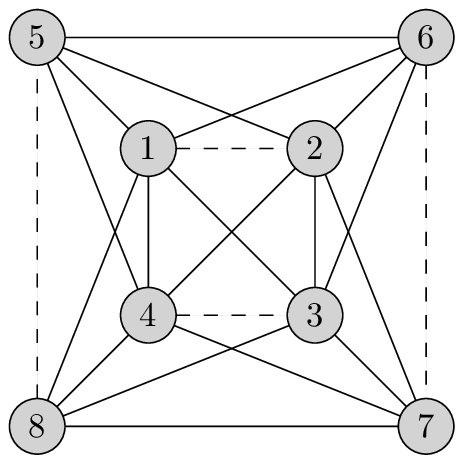}}
  \end{minipage}
  \hfill
  \begin{minipage}[b]{.37\textwidth}
     \centering
     \subfloat[\label{fig:nonrac-graph-planar-part}{}]
     {\includegraphics[width=.7\textwidth]{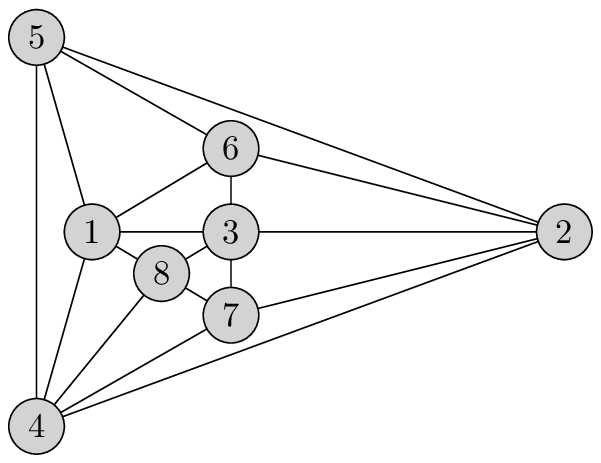}}
  \end{minipage}
  \caption{(a) A graph with $8$ vertices and $22$ edges which does not admit a RAC drawing\
  \cite{EL11}. (b) A decomposition of the
  graph of Fig.\ref{fig:nonrac-graph} into a planar graph (solid edges; a planar drawing is
  given in Fig.\ref{fig:nonrac-graph-planar-part}) and a matching (dashed edges),
  which implies that a planar graph and a matching do not always admit a $\GSimRAC$ drawing; their union is not a RAC graph.}
  \label{fig:nonrac-graph-example}
\end{figure}

If two graphs do not admit a geometric simultaneous drawing they,
obviously, do not admit a $\GSimRAC$ drawing. For instance, since it
is known that there exists a planar graph and a matching that do not
admin a geometric simultaneous drawing \cite{CvKLMSV11}, as a
consequence, the same graph and matching do not admit a $\GSimRAC$
drawing either. Figure~\ref{fig:nonrac-graph-example} depicts an
alternative and simpler technique to prove such negative results,
which is based on the fact that not all graphs that obey the
RAC-size constraint are eventually RAC graphs. On the other hand, as
we will shortly see, if two graphs always admit a geometric
simultaneous drawing, it is not necessary that they also admit a
$\GSimRAC$ drawing.

The $\GSimRAC$ drawing problem is of
interest since it combines two current research topics in graph
drawing. Our motivation to study this
problem rests on the work of Didimo et al.\ \cite{DEL09} who proved
that the crossing graph of a geometric RAC drawing is
bipartite\footnote{This can be interpreted as follows: ``If two
edges of a geometric RAC drawing cross a third one, then these two
edges must be parallel.''}. Thus, the edges of a geometric RAC
drawing of a graph $G=(V,E)$ can be partitioned into two sets $E_1$
and $E_2$, such that no two edges of the same set cross. So, the
problem we study is, in a sense, equivalent to the problem of
finding a geometric RAC drawing of an input graph (if one exists),
given its crossing graph.

\section{Related Work and our Results}
\label{sec:related-work}

Didimo et al.\ \cite{DEL09} were the first to study the geometric
RAC drawing problem and proved that any graph with $n\geq 3$
vertices that admits a geometric RAC drawing has at most $4n-10$
edges. Arikushi et al.\ \cite{AFKMT10} presented bounds on the
number of edges of polyline RAC drawings with at most one or two
bends per edge. Angelini et al.\ \cite{ACBDFKS09} presented acyclic
planar digraphs that do not admit upward geometric RAC drawings and
proved that the corresponding decision problem is $\NP$-hard.
Argyriou et al.\ \cite{ABS11} proved that it is $\NP$-hard to decide
whether a given graph admits a geometric RAC drawing (i.e., the
upwardness requirement is relaxed). Di Giacomo et al.\
\cite{DGDLM10} presented tradeoffs on the maximum number of bends
per edge, the required area and the crossing angle resolution.
Didimo et al.\ \cite{DEL10} characterized classes of complete
bipartite graphs that admit geometric RAC drawings. Van Kreveld
\cite{vK10} showed that the quality of a planar drawing of a planar
graph (measured in terms of area required, edge-length and angular
resolution) can be improved if one allows right-angle crossings.
Eades and Liotta\ \cite{EL11} proved that a \emph{maximally dense
RAC graph} (i.e., $|E|=4|V|-10$) is also $1$-planar, i.e., it admits
a drawing in which every edge is crossed at most once.

Regarding the geometric simultaneous graph drawing problem, Brass et
al.\ \cite{BCDE07} presented algorithms for drawing simultaneously
(a)~two paths, (b)~two cycles and, (c)~two caterpillars.
Estrella-Balderrama et al.\ \cite{EGJPSS07}
proved that the problem of determining whether two planar graphs
admit a geometric simultaneous drawing is $\NP$-hard. Erten and
Kobourov\ \cite{EK05} showed that a planar graph and a path cannot
always be drawn simultaneously.
Geyer,
Kaufmann and Vrt'o\ \cite{GKV09}, showed that a geometric
simultaneous drawing of two trees does not always exist. Angelini et
al.\ \cite{AGKN10} proved the same result for a path and a tree.
Cabello et al.\ \cite{CvKLMSV11} showed that a geometric
simultaneous drawing of a matching and (a)~a wheel, (b)~an outerpath
or, (c)~a tree always exists, while there exist a planar graph and a
matching that cannot be drawn simultaneously. For a quick overview
of known results
refer to Table~$1$ of
\cite{FKK09}.

A closely related problem to the $\GSimRAC$ drawing problem is the
following: \emph{Given a planar embedded graph $G$, determine a
geometric drawing of $G$ and its dual $G^*$ (without the face-vertex
corresponding to the external face) such that: (i)~$G$ and $G^*$ are
drawn planar, (ii)~each vertex of the dual is drawn inside its
corresponding face of $G$ and, (iii)~the primal-dual edge crossings
form right-angles}. We refer to this problem as the \emph{geometric
simultaneous Graph-Dual RAC drawing problem} (or
\emph{GDual-GSimRAC} for short). Brightwell and
Scheinermann~\cite{BS93} proved that the $\GDual$ drawing problem
always admits a solution if the input graph is a triconnected planar
graph. To the best of our knowledge, this is the only result which
incorporates the requirement that the primal-dual edge crossings
form right-angles. Erten and Kobourov~\cite{EK05}, presented an
$O(n)$ time algorithm that results into a simultaneous drawing but,
unfortunately, not a RAC drawing of a triconnected planar graph and
its dual on an $O(n^2)$
grid, where $n$ is the number of vertices of $G$ and $G^*$.

This paper is structured as follows: In
Section~\ref{sec:wheel-cycle}, we demonstrate that if two graphs
always admit a geometric simultaneous drawing, it is not necessary
that they also admit a $\GSimRAC$ drawing. In
Section~\ref{sec:cycle-matching}, we prove that a cycle and a
matching always admit a $\GSimRAC$ drawing. In
Section~\ref{sec:planar-dual}, we examine variations of the $\GDual$
drawing problem. We conclude in Section~\ref{sec:conclusion} with
open problems.

Before we proceed with the description of our results, we introduce
some necessary notation. Let $G=(V,E)$ be a simple, undirected graph
drawn on the plane. We denote by $\Gamma(G)$ the drawing of $G$. By
$x(v)$ and $y(v)$, we denote the $x$- and $y$-coordinate of $v \in
V$ in $\Gamma(G)$. We refer to the vertex (edge) set of $G$ as
$V(G)$ ($E(G)$). Given two graphs $G$ and $G'$, we denote by $G \cup
G'$ the graph induced by the union of $G$ and $G'$.

\section{A Wheel and a Cycle: A Negative Result}
\label{sec:wheel-cycle}

In this section, we demonstrate that if two graphs always admit a
geometric simultaneous drawing, it is not necessary that they also
admit a $\GSimRAC$ drawing. We achieve this by showing that there
exists a wheel and a cycle which do not admin a $\GSimRAC$ drawing.
Cabello et al.\ \cite{CvKLMSV11} have shown that a geometric
simultaneous drawing of a wheel and a cycle always exists.

\begin{figure}[h!tb]
  \centering
  \begin{minipage}[b]{.32\textwidth}
     \raggedleft
     \subfloat[\label{fig:ata-graph-1}{}]
     {\includegraphics[width=\textwidth]{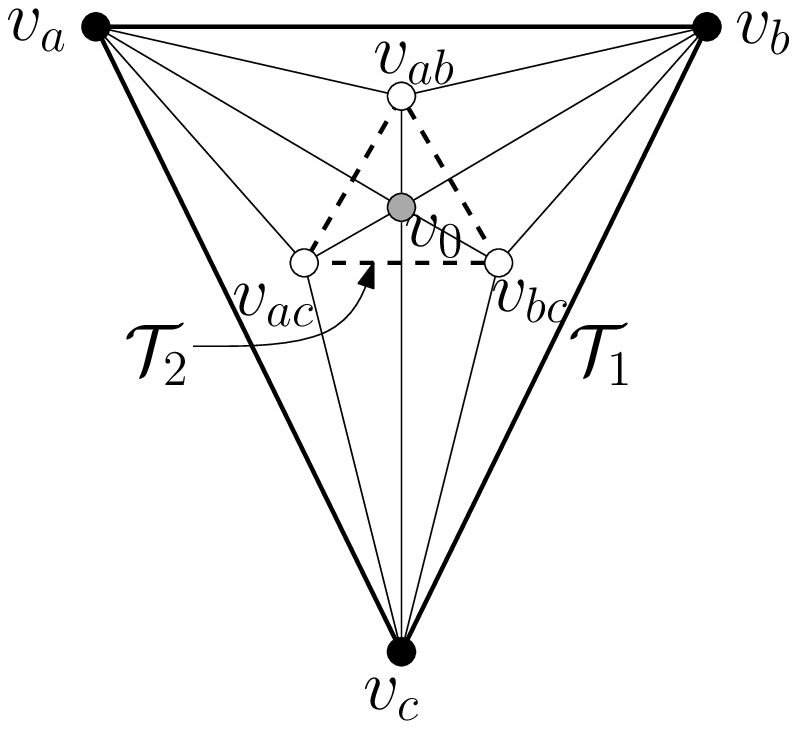}}
  \end{minipage}
  \hfill
  \begin{minipage}[b]{.32\textwidth}
     \raggedright
     \subfloat[\label{fig:ata-graph-2}{}]
     {\includegraphics[width=\textwidth]{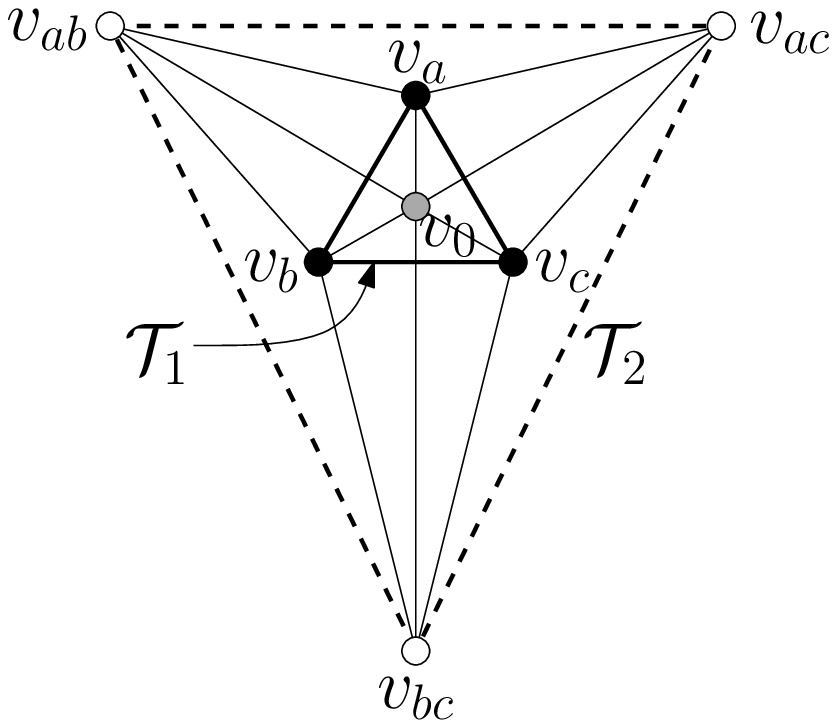}}
  \end{minipage}
  \hfill
  \begin{minipage}[b]{.32\textwidth}
     \raggedright
     \subfloat[\label{fig:ata-graph-decomposition}{}]
     {\includegraphics[width=\textwidth]{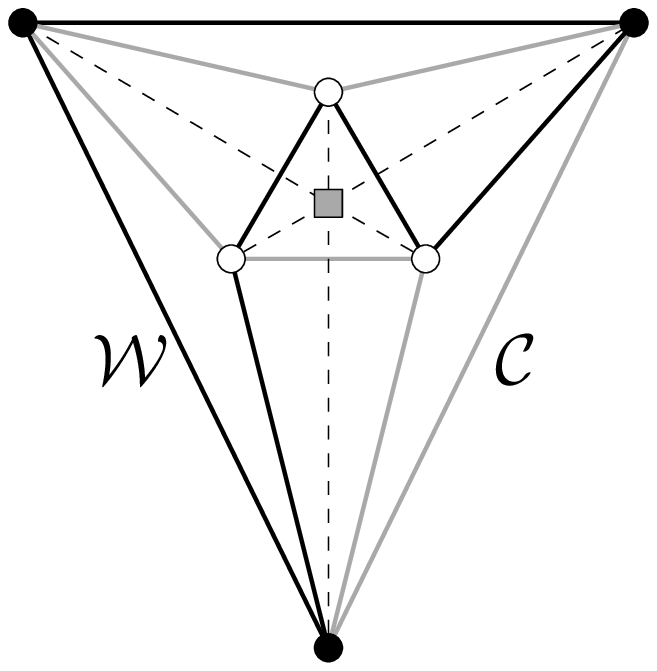}}
  \end{minipage}
  \caption{(a)-(b)~Two different RAC drawings of the augmented triangle antiprism graph with different combinatorial embeddings.
  (c)~The union of wheel $\mathcal{W}$ (solid and dashed black edges) and cycle $\mathcal{C}$ (gray edges) is
  the augmented triangle antiprism graph.}
  \label{fig:ata-graph}
\end{figure}

Our proof utilizes the \emph{augmented triangle antiprism graph}
\cite{ABS11,DEL09}, depicted in Figure~\ref{fig:ata-graph-1}. The
augmented triangle antiprism graph contains two triangles
$\mathcal{T}_1$ and $\mathcal{T}_2$ (refer to the dashed and bold
drawn triangles in Figure~\ref{fig:ata-graph-1}) and a ``central''
vertex $v_0$ incident to the vertices of $\mathcal{T}_1$ and
$\mathcal{T}_2$. If we delete the central vertex, the remaining
graph corresponds to the skeleton of a triangle antiprism and it is
commonly referred to as \emph{triangle antiprism graph}. Didimo et
al.\ \cite{DEL09} used the augmented triangle antiprism graph as an
example of a maximally dense RAC graph (i.e., $|E|=4|V|-10$).

\begin{lemma}
\label{lem:ata-graph} The geometric RAC drawings of the augmented
triangle antiprism graph define exactly two combinatorial
embeddings.
\end{lemma}

\begin{sketchofproof}
Figures~\ref{fig:ata-graph-1} and~\ref{fig:ata-graph-2} illustrate
this property. The proof of the lemma is based on the following
properties:

\begin{enumerate}[i)]
\item In any RAC drawing of the augmented triangle antiprism graph,
triangles $\mathcal{T}_1$ and $\mathcal{T}_2$ do not cross.

\item In any RAC drawing of the augmented triangle antiprism graph,
the external face is bounded by three edges~\cite{EL11}.

\item There does not exist a RAC drawing of the augmented triangle
antiprism graph in which the $3$-cycle incident to the external face
consists of two vertices of $\mathcal{T}_1$ ($\mathcal{T}_2$) and
one vertex of $\mathcal{T}_2$ ($\mathcal{T}_1$).

\item There does not exist a RAC drawing of the augmented triangle
antiprism graph in which the $3$-cycle incident to the external face
consists of the central vertex $v_0$ and two vertices of either
$\mathcal{T}_1$ or two vertices of $\mathcal{T}_2$.

\item There does not exist a RAC drawing of the augmented triangle
antiprism graph in which the $3$-cycle incident to the external face
consists of the central vertex $v_0$, one vertex of $\mathcal{T}_1$
and one vertex of $\mathcal{T}_2$.

\item In any RAC drawing of the augmented triangle antiprism
graph, the central vertex $v_0$ lies in the interior of both
$\mathcal{T}_1$ and $\mathcal{T}_2$.
\end{enumerate}

Due to space constraints, we omit the detailed proofs of these
properties. The proofs make use of elementary geometric properties,
they heavily use Lemma~$2$ of~\cite{DEL09} and Property~$2$
of~\cite{ACBDFKS09}, and are based on an exhaustive cases analysis
on the relative positions of (a)~the central vertex $v_0$, and,
(b)~triangles $\mathcal{T}_1$ and $\mathcal{T}_2$.
\end{sketchofproof}

\begin{theorem}
\label{thm:wheel-cycle} There exists a wheel and a cycle which do
not admin a $\GSimRAC$ drawing.
\end{theorem}
\begin{proof}
We denote the wheel by $\mathcal{W}$ and the cycle by $\mathcal{C}$.
The counterexample is depicted in
Figure~\ref{fig:ata-graph-decomposition}. The center of
$\mathcal{W}$ is marked by a box, the spokes of $\mathcal{W}$ are
drawn as dashed line-segments, while the rim of $\mathcal{W}$ is
drawn in bold. Cycle $\mathcal{C}$ is drawn in gray. The graph
induced by the union of $\mathcal{W}$ and $\mathcal{C}$ (which in a
$\GSimRAC$ drawing of $\mathcal{W}$ and $\mathcal{C}$ should be
drawn with right-angle crossings) is the augmented triangle
antiprism graph, which, by Lemma~\ref{lem:ata-graph}, has exactly
two RAC combinatorial embeddings. However, in none of them wheel
$\mathcal{W}$ can be drawn planar. This completes the proof.\qed
\end{proof}

\section{A Cycle and a Matching: A Positive Result}
\label{sec:cycle-matching}

\begin{figure}[h!tb]
  \centering
  \includegraphics[width=.8\textwidth]{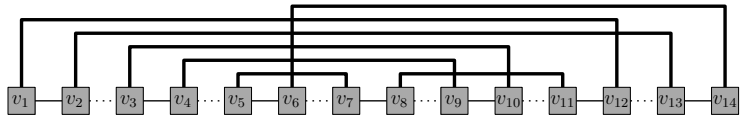}
  \caption{An example of a path $\mathcal{P}$ and a matching $\mathcal{M}$.
  The path appears at the bottom of the figure. The edges of $\mathcal{M}$ are drawn bold, with two bends each.
  The edges of path $\mathcal{P}$ form two matchings, i.e., $\mathcal{P}_{odd}$
  and $\mathcal{P}-\mathcal{P}_{odd}$. The edges of $\mathcal{P}_{odd}$ are drawn solid, while the edges of $\mathcal{P}-\mathcal{P}_{odd}$ dotted. }
  \label{fig:path-matching-sample}
\end{figure}

In this section, we first prove that a path and a matching always
admit a $\GSimRAC$ drawing and then we show that a cycle and a
matching always admit a $\GSimRAC$ drawing as well. Note that the
union of a path and a matching is not necessarily a planar graph.
Cabello et al.\ \cite{CvKLMSV11} provide an example of a path and a
matching, which form a subdivision of $K_{3,3}$. We denote the path
by $\mathcal{P}$ and the matching by $\mathcal{M}$. Let $v_1
\rightarrow v_2 \rightarrow \ldots \rightarrow v_n$ be the edges of
$\mathcal{P}$ (see Figure~\ref{fig:path-matching-sample}). In order
to keep the description of our algorithm simple, we will initially
assume that $n$ is even and $|E(\mathcal{M})|=n/2$. Later on this
section, we will describe how to cope with the cases where $n$ is
odd or $|E(\mathcal{M})|<n/2$. Recall that by the definition of the
$\GSimRAC$ drawing problem, $\mathcal{P}$ and $\mathcal{M}$ do not
share an edge, i.e., $E(\mathcal{P}) \cap E(\mathcal{M}) =
\emptyset$.

The basic idea of our algorithm is to identify in the graph induced
by the union of $\mathcal{P}$ and $\mathcal{M}$ a set of cycles
$\mathcal{C}_1, \mathcal{C}_2, \ldots, \mathcal{C}_k$, $k \leq n/4$,
such that: (i)~$|E(\mathcal{C}_1)| + |E(\mathcal{C}_2)| + \ldots +
|E(\mathcal{C}_k)|=n$, (ii)~$\mathcal{M} \subseteq \mathcal{C}_1
\cup \mathcal{C}_2 \cup \ldots \cup \mathcal{C}_k$, and, (iii)~the
edges of cycle $\mathcal{C}_i$,  $i=1,2,\ldots,k$ alternate between
edges of $\mathcal{P}$ and $\mathcal{M}$. Note that properties (i)
and (ii) imply that the cycle collection will contain half of
$\mathcal{P}$'s edges and all of $\mathcal{M}$'s edges. In our
drawing, these edges will not cross with each other. The remaining
edges of $\mathcal{P}$ will introduce only right-angle crossings
with the edges of $\mathcal{M}$.

\begin{figure}[t!hb]
  \centering
  \includegraphics[width=.67\textwidth]{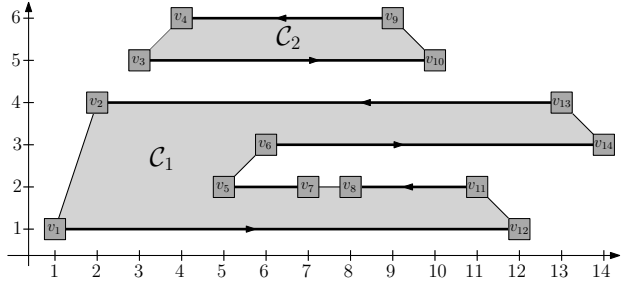}
  \caption{$\mathcal{P}_{odd} \cup \mathcal{M}$ (of Fig.\ref{fig:path-matching-sample})
  consists of cycles $\mathcal{C}_1$ and $\mathcal{C}_2$. The edges of $\mathcal{P}_{odd}$
  are drawn solid, while the edges of $\mathcal{M}$ are drawn bold.}
  \label{fig:path-matching-cycle-decomposition}
\end{figure}

Let $\mathcal{P}_{odd}$ be a subgraph of $\mathcal{P}$ which
contains each second edge of $\mathcal{P}$, starting from its first
edge, i.e., $E(\mathcal{P}_{odd}) = \{(v_i,v_{i+1});~1\leq i <
n,~i\text{ is odd}\}$. In Figure~\ref{fig:path-matching-sample}, the
edges of $\mathcal{P}_{odd}$ are drawn solid. Clearly,
$\mathcal{P}_{odd}$ is a matching. Since we have assumed that $n$ is
even, $\mathcal{P}_{odd}$ contains exactly $n/2$ edges. Hence,
$|E(\mathcal{P}_{odd})|=|E(\mathcal{M})|$. In addition,
$\mathcal{P}_{odd}$ covers all vertices of $\mathcal{P}$, and,
$E(\mathcal{P}_{odd}) \cap E(\mathcal{M}) = \emptyset$. The later
equation trivially follows from our initial hypothesis, which states
that $E(\mathcal{P}) \cap E(\mathcal{M}) = \emptyset$. We conclude
that $\mathcal{P}_{odd} \cup \mathcal{M}$ is a $2$-regular graph.
Thus, each connected component of $\mathcal{P}_{odd} \cup
\mathcal{M}$ corresponds to a cycle of even length, which alternates
between edges of $\mathcal{P}_{odd}$ and $\mathcal{M}$. This is the
cycle collection mentioned above (see
Figure~\ref{fig:path-matching-cycle-decomposition}).

Initially, we fix the $x$-coordinate of each vertex of $\mathcal{P}$
by setting $x(v_i) = i$, $1 \leq i \leq n$. This ensures that
$\mathcal{P}$ is $x$-monotone and hence planar. Later on, we will
slightly change the $x$-coordinate of some vertices of $\mathcal{P}$
(without affecting $\mathcal{P}$'s monotonicity). The $y$-coordinate
of each vertex of $\mathcal{P}$, is established by considering the
cycles of $\mathcal{P}_{odd} \cup \mathcal{M}$\footnote{The
algorithm can be adjusted so that the $x$ and $y$ coordinates of
each vertex are computed at the same time. We have chosen to compute
them separately in order to simplify the presentation.}.

We draw each of these cycles in turn. More precisely, assume that
zero or more cycles have been completely drawn and let $\mathcal{C}$
be the cycle in the cycle collection which contains the leftmost
vertex, say $v_i$, of $\mathcal{P}$ that has not been drawn yet
(initially, $v_i$ is identified by $v_1$). Then, vertex $v_i$ should
be an odd-indexed vertex and thus $(v_i,v_{i+1})$ belongs in
$\mathcal{C}$. Orient cycle $\mathcal{C}$ so that vertex $v_{i}$ is
the first vertex of cycle $\mathcal{C}$ and $v_{i+1}$ is the last
(see Figure~\ref{fig:path-matching-cycle-decomposition}). Based on
this orientation, we will draw the edges of $\mathcal{C}$ in a
snake-like fashion, starting from vertex $v_i$ and reaching vertex
$v_{i+1}$ last. The first edge to be drawn is incident to vertex
$v_i$ and belongs to $\mathcal{M}$. We draw it as a horizontal
line-segment at the bottommost available layer in the produced
drawing (initially, $L_1:y=1$). Since cycle $\mathcal{C}$ alternates
between edges of $\mathcal{P}_{odd}$ and $\mathcal{M}$, the next
edge to be drawn belongs to $\mathcal{P}_{odd}$ followed by an edge
of $\mathcal{M}$. If we can draw both of them in the current layer
without introducing edge overlaps, we do so. Otherwise, we employ an
additional layer. We continue in the same manner, until edge
$(v_i,v_{i+1})$ is reached in the traversal of cycle $\mathcal{C}$.
This edge connects two consecutive vertices of $\mathcal{P}$ that
are the leftmost in the drawing of $\mathcal{C}$. Therefore, edge
$(v_i,v_{i+1})$ can be added in the drawing of $\mathcal{C}$ without
introducing any crossings. Thus, cycle $\mathcal{C}$ is drawn
planar.

So far, we have drawn all edges of $\mathcal{M}$ and half of the
edges of $\mathcal{P}$ (i.e., $\mathcal{P}_{odd}$) and we have
obtained a planar drawing in which all edges of $\mathcal{M}$ are
drawn as horizontal, non-overlapping line segments. In the worst
case, this drawing occupies $n/2$ layers.

\begin{figure}[t!]
  \begin{minipage}[b]{.67\textwidth}
     \raggedleft
     \subfloat[\label{fig:path-matching-addon-edges}{A drawing obtained by incorporating the edges of $\mathcal{P}-\mathcal{P}_{odd}$
  into the drawing of Fig.\ref{fig:path-matching-cycle-decomposition}.}]
     {\includegraphics[width=\textwidth]{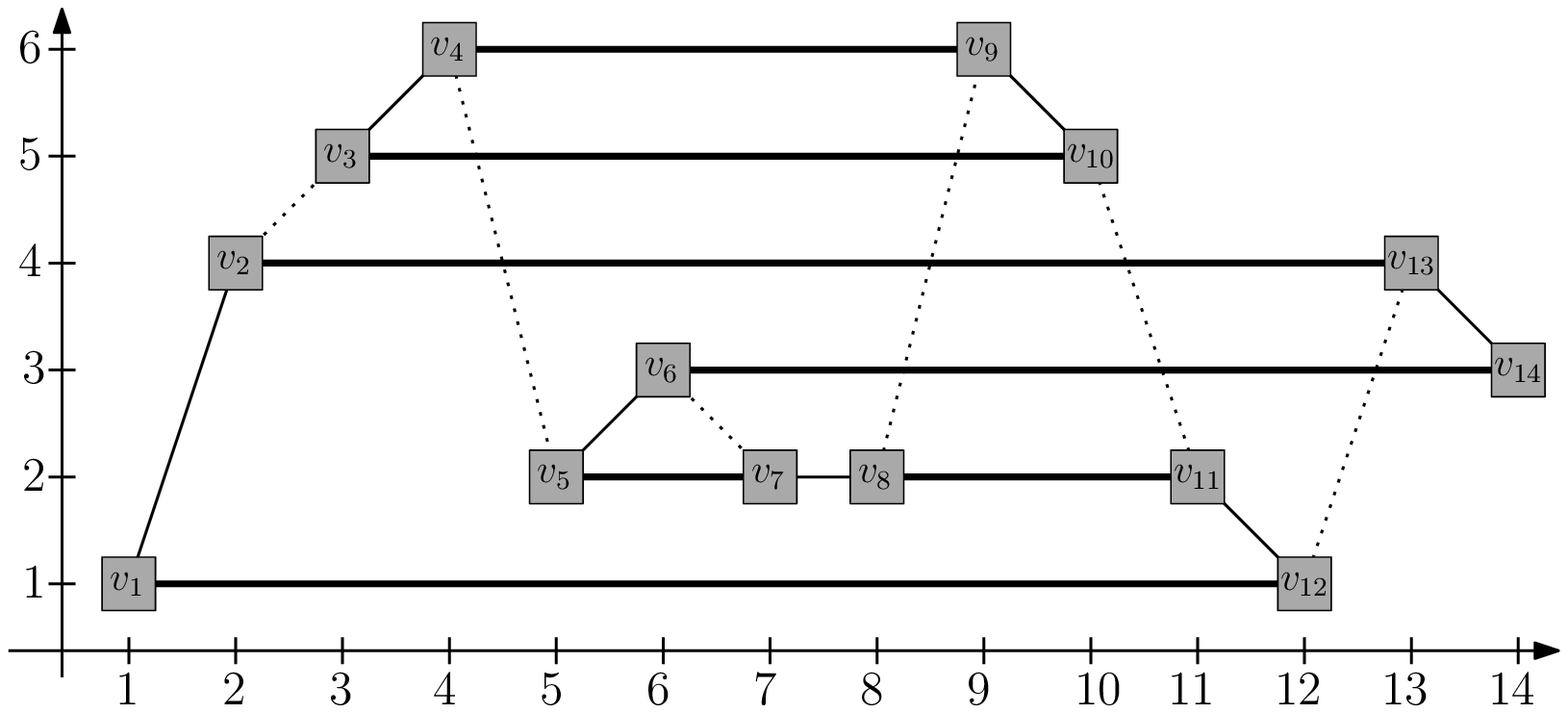}}
  \end{minipage}
  \hfill
  \begin{minipage}[b]{.32\textwidth}
     ~
  \end{minipage}
  \vfill
  \begin{minipage}[b]{.67\textwidth}
     \raggedleft
     \subfloat[\label{fig:path-matching-final}{A drawing obtained by moving the even-indexed vertices of
  $\mathcal{P}$ in the drawing of Fig.\ref{fig:path-matching-addon-edges} one unit to the right.}]
     {\includegraphics[width=\textwidth]{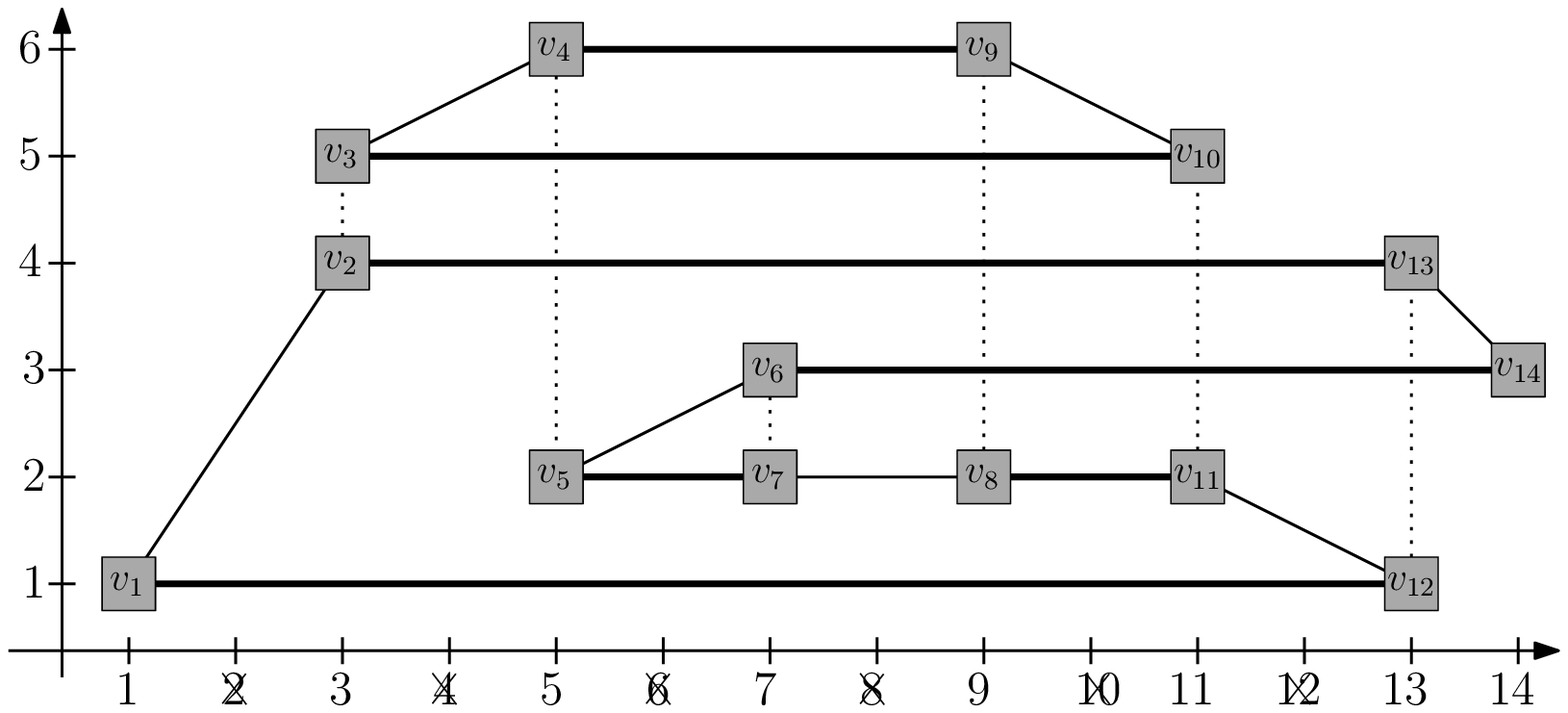}}
  \end{minipage}
  \hfill
  \begin{minipage}[b]{.32\textwidth}
     \raggedright
     \subfloat[\label{fig:path-matching-final-improved}{A compact $\GSimRAC$ drawing}]
     {\includegraphics[width=\textwidth]{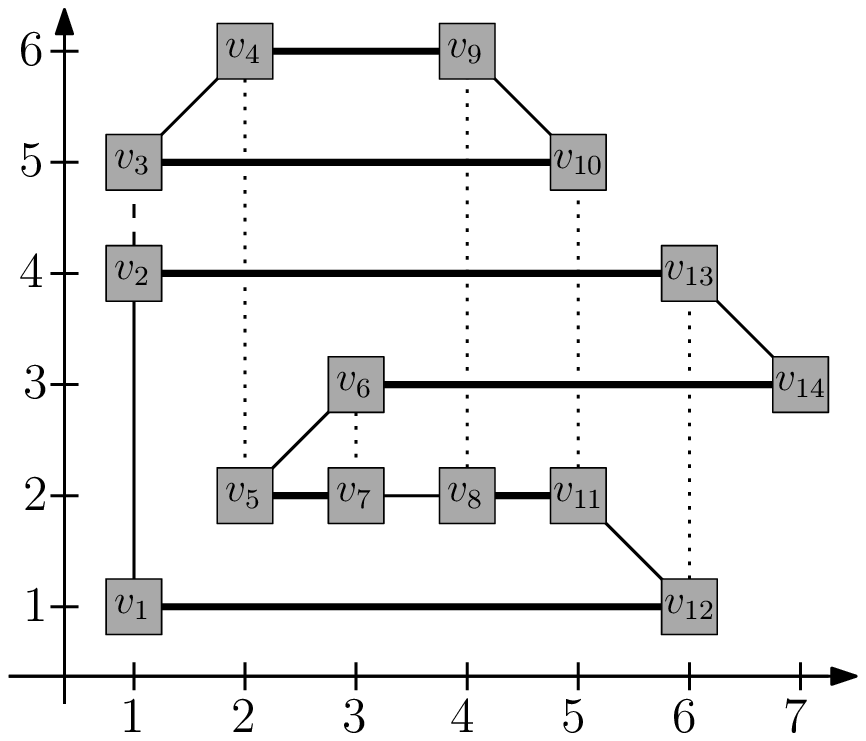}}
  \end{minipage}
  \caption{In all drawings, the edges of $\mathcal{P}_{odd}$ are drawn solid, while the edges of $\mathcal{P}-\mathcal{P}_{odd}$
  dotted. The edges of $\mathcal{M}$ are drawn bold.}
  \label{fig:path-matching-final-drawing}
\end{figure}

We proceed to incorporate the remaining edges of $\mathcal{P}$, i.e,
the ones that belong in $\mathcal{P}-\mathcal{P}_{odd}$, into the
drawing (refer to the dotted drawn edges of
Figure~\ref{fig:path-matching-addon-edges}). Since $x(v_i) = i$,
$i=1,2,\ldots,n$, the edges of $\mathcal{P}$ do not cross with each
other and therefore $\mathcal{P}$ is drawn planar. In contrast, an
edge of $\mathcal{P}-\mathcal{P}_{odd}$ may cross multiple edges of
$\mathcal{M}$, and, these crossings do not form right-angles (see
Figure~\ref{fig:path-matching-addon-edges}). However, it is not
difficult to fix this. A simple approach suggests to move each
even-indexed vertex of $\mathcal{P}$ one unit to the right (keeping
its $y$-coordinate unchanged), expect from the last vertex of
$\mathcal{P}$. Then, the endpoints of the edges of
$\mathcal{P}-\mathcal{P}_{odd}$ have exactly the same $x$-coordinate
and cross at right-angles the edges of $\mathcal{M}$ which are drawn
as horizontal line-segments. The path remains $x$-monotone (but not
strictly anymore) and hence planar. In addition, it is not possible
to introduce vertex overlaps, since in the produced drawing each
edge of $\mathcal{M}$ has at least two units length (recall that
$E(\mathcal{P}) \cap E(\mathcal{M}) = \emptyset$). Since the
vertices of the drawing do not occupy even $x$-coordinates, the
width of the drawing can be reduced from $n$ to $n/2+1$ (see
Figure~\ref{fig:path-matching-final}). We can further reduce the
width of the produced drawing by merging consecutive columns that do
not interfere in $y$-direction into a common column (see
Figure~\ref{fig:path-matching-final-improved}). However, this
post-processing does not result into a drawing of asymptotically
smaller area.

In order to complete the description of our algorithm, it remains to
consider the cases where $n$ is odd or $|E(\mathcal{M})|<n/2$. Both
cases can be treated similarly. If $n$ is odd or
$|E(\mathcal{M})|<n/2$, there exist vertices of $\mathcal{P}$ which
are not covered by matching $\mathcal{M}$. As long as there exist
such vertices, we can momentarily remove them from the path by
contracting each subpath consisting of degree-$2$ vertices into a
single edge. By this procedure, we obtain a new path $\mathcal{P}'$,
so that $\mathcal{M}$ covers all vertices of $\mathcal{P}'$. If we
draw $\mathcal{P}'$ and $\mathcal{M}$ simultaneously, then it is
easy to incorporate the removed vertices in the produced drawing,
since they do not participate in $\mathcal{M}$. The following
theorem summarizes our result.

\begin{theorem}
\label{thm:path-matching} A path and a matching always admit a
$\GSimRAC$ drawing on an $(n/2+1) \times n/2$ integer grid.
Moreover, the drawing can be computed in linear time.
\end{theorem}
\begin{proof}
Finding the cycles of $\mathcal{P}_{odd} \cup \mathcal{M}$ can be
easily done in $O(n)$ time, where $n$ is the number of vertices of
$\mathcal{P}$. We simply identify the leftmost vertex of each cycle
and then we traverse it. Having computed the cycle collection of
$\mathcal{P}_{odd} \cup \mathcal{M}$, the coordinates of the
vertices are computed in $O(n)$ total time by a simple traversal of
the cycles. \qed
\end{proof}

\begin{figure}[h!tb]
  \centering
  \includegraphics[width=.6\textwidth]{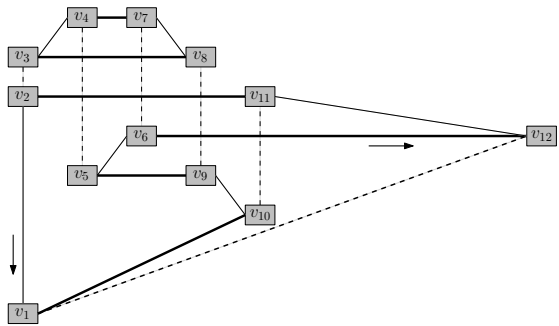}
  \caption{A $\GSimRAC$ drawing of a cycle and a matching.}
  \label{fig:cycle-matching}
\end{figure}

We extend the algorithm that produces a $\GSimRAC$ drawing of a path
and a matching to also cover the case of a cycle $\mathcal{C}$ and a
matching $\mathcal{M}$. The idea is quite simple (see
Figure~\ref{fig:cycle-matching}). If we remove an edge from the
input cycle, the remaining graph is a path $\mathcal{P}$. Then, we
apply the developed algorithm and obtain a $\GSimRAC$ drawing of
$\mathcal{P}$ and $\mathcal{M}$, in which the first vertex of
$\mathcal{P}$ is drawn at the bottommost layer (hence its incident
edge in $\mathcal{M}$ is not crossed), and, the last vertex of
$\mathcal{P}$ is drawn rightmost. With these two properties, it is
not difficult to add the removed edge, between the first and the
last vertex of $\mathcal{P}$. Simply move the first vertex of
$\mathcal{P}$ at most $n/2+2$ units downwards (keeping its
$x$-coordinate unchanged) and the last vertex of $\mathcal{P}$ at
most $n/2+1$ units rightwards (keeping its $y$-coordinate
unchanged). Then, the insertion in the drawing of the edge that
closes the cycle does not introduce any crossings.

\begin{theorem}
\label{thm:cycle-matching} A cycle and a matching always admit a
$\GSimRAC$ drawing on an $(n+2)\times(n+2)$ integer grid. Moreover,
the drawing can be computed in linear time.
\end{theorem}

\begin{corollary}
\label{cor:cycle-matching} Let $G$ be a simple connected graph that
can be decomposed into a matching and either a hamiltonian path or a
hamiltonian cycle. Then, $G$ is a RAC graph.
\end{corollary}

\section{A Planar Graph and its Dual: An Interesting Variation}
\label{sec:planar-dual}

In this section, we examine the  $\GDual$ drawing problem. This
problem can be considered as a variation of the $\GSimRAC$ drawing
problem, where the first graph (i.e., the planar graph) determines
the second one (i.e., the dual) and places restrictions on its
layout. Recall that according to the $\GDual$ drawing problem, we
are given a planar embedded graph $G$ and the main task is to
determine a geometric drawing of $G$ and its dual $G^*$ (without the
face-vertex corresponding to the external face) such that: (i)~$G$
and $G^*$ are drawn planar, (ii)~each vertex of the dual is drawn
inside its corresponding face of $G$ and, (iii)~the primal-dual edge
crossings form right-angles. As already stated in
Section~\ref{sec:related-work}, Brightwell and
Scheinermann~\cite{BS93} proved that this is always feasible if the
input graph is a triconnected planar graph. For the general case of
planar graphs, we demonstrate by an example that it is not always
possible to compute such a drawing, and thus, we concentrate our
study in the case of outerplanar graphs.

\begin{figure}[t!hb]
  \centering
  \begin{minipage}[b]{.43\textwidth}
     \centering
     \subfloat[\label{fig:fixeddrawing}{}]
     {\includegraphics[width=.9\textwidth]{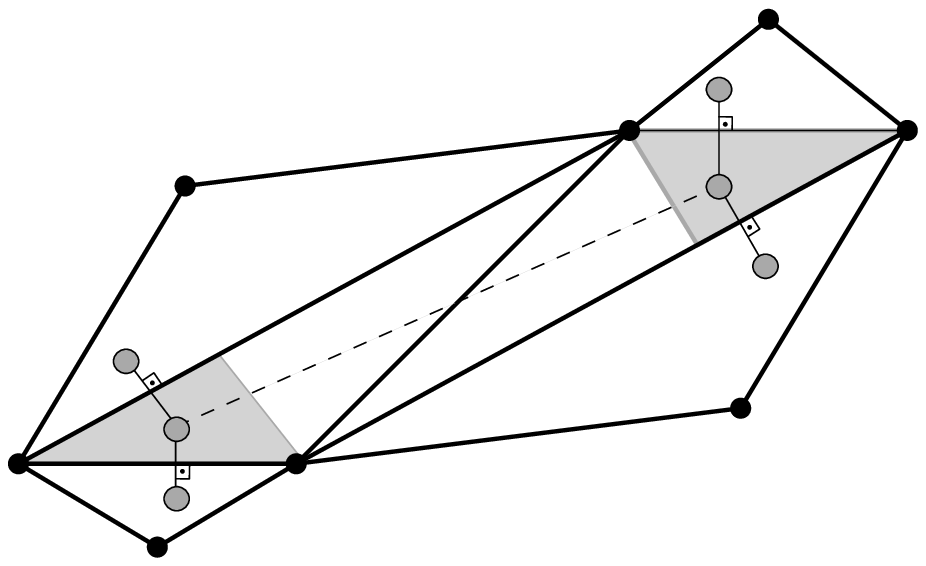}}
  \end{minipage}
  \hfill
  \begin{minipage}[b]{.52\textwidth}
     \centering
     \subfloat[\label{fig:planar-graph-and-dual}{}]
     {\includegraphics[width=.8\textwidth]{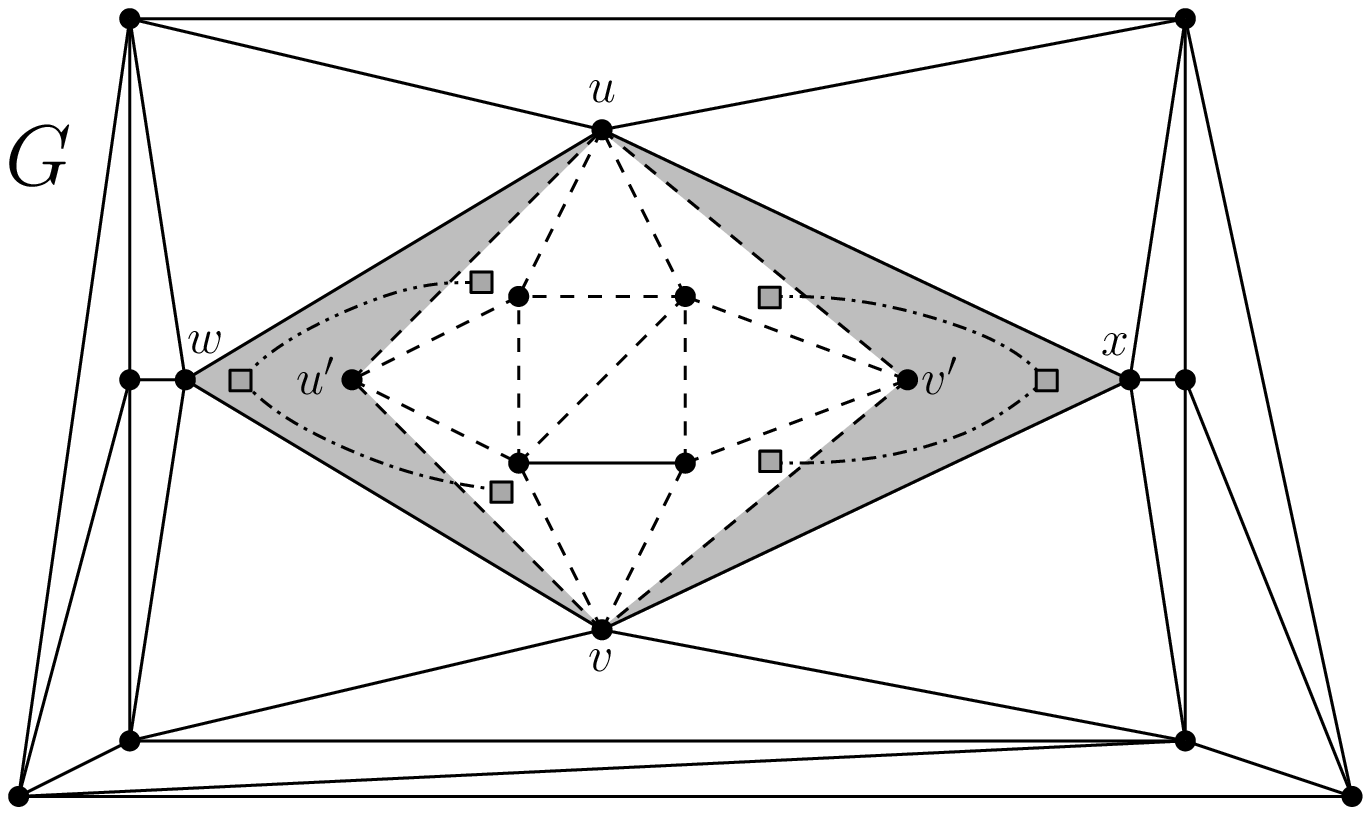}}
  \end{minipage}
  \caption{(a)~The input planar drawing of the primal graph $G$ is sketched with black colored vertices and bold edges and should remain
  unchanged in the output. The vertices of the dual $G^*$ are colored gray. Then, the dual's dashed drawn edge will inevitably introduce a non right-angle crossing.
  (b)~An example of a planar graph $G$ for which it is not feasible to determine a geometric drawing of $G$ and its dual $G^*$, such that $G$ and $G^*$
are drawn planar and the primal-dual edge crossings form
right-angles. The problematic faces are drawn in gray.}
  \label{fig:ata-graph}
\end{figure}

Initially, we consider the case where the planar drawing $\Gamma(G)$
of graph $G$ is specified as part of the input and it is required
that it remains unchanged in the output, we demonstrate by an
example that it is not always feasible to incorporate $G^*$ into
drawing $\Gamma(G)$ and obtain a $\GDual$ drawing of $G$ and $G^*$.
The example is illustrated in Figure~\ref{fig:fixeddrawing}.

In the following, we prove that if the input graph is a planar
embedded graph, then the $\GDual$ drawing problem does not always
admit a solution.

\begin{theorem}
\label{thm:planar-embedded}Given a planar embedded graph $G$, a
$\GDual$ drawing of $G$ and its dual $G^*$ does not always exist.
\end{theorem}

\begin{proof}
The planar graph $G$ used to establish the theorem is depicted in
Figure~\ref{fig:planar-graph-and-dual}, where the vertices drawn as
boxes belong to the dual graph $G^*$. Observe that the subgraph
drawn with dashed edges is a triconnected planar graph. Thus, it has
a unique planar embedding (up to a reflexion). If we replace this
subgraph by an edge, the remaining primal graph, is also
triconnected. Therefore, the graph of our example is a subdivision
of a triconnected graph and, thus, it has two planar combinatorial
embeddings obtained by reflections of the triconnected planar
subgraph, at vertices $u$ and $v$, i.e., either vertex $u'$ is to
the ``left'' of $v'$, or to its ``right''. Now, observe that the
dual graph should have two vertices within the gray-colored faces of
Figure~\ref{fig:planar-graph-and-dual} (refer to the vertices which
are drawn as boxes). Each of these two vertices is incident to two
vertices of the dual that lie within the triangular faces of the
dashed drawn subgraph of $G$, incident to the two gray-colored
faces. We observe that in order to have a RAC drawing of both $G$
and $G^*$ both quadrilaterals $uu'vw$ and $uv'vx$ must be drawn
convex, which is impossible. \qed
\end{proof}

\begin{theorem}
\label{thm:outerplanar-dual}Given an outerplane embedding of an
outerplanar graph $G$, it is always feasible to determine a $\GDual$
drawing of $G$ and its dual $G^*$.
\end{theorem}

\begin{proof}
The proof is given by a recursive geometric construction which
computes a $\GDual$ drawing of $G$ and its dual. Consider an
arbitrary edge $(u,v)$ of the outerplanar graph that does not belong
to its external face and let $f$ and $g$ be the faces to its left
and the right side, respectively, as we move along $(u,v)$ from
vertex $u$ to vertex $v$. Then, $(f,g)$ is an edge of the dual graph
$G^*$. Since the dual of an outerplanar graph is a tree, the removal
of edge $(f,g)$ results in two trees $T_f$ and $T_g$ that can be
considered to be rooted at vertices $f$ and $g$ of $G^*$,
respectively. For the recursive step of our drawing algorithm, we
assume that we have already produced a $\GDual$ drawing for $T_f$
and its corresponding subgraph of $G$ that satisfies the following
invariant properties:
\begin{description}
\item{\textbf{I-P1}:} \emph{Edge $(u,v)$ is drawn on the external face of the GDual-GSimRAC drawing constructed so
far.} Let $u$ and $v$ be drawn at points $p_u$ and $p_v$,
respectively. Denote by $\ell_{u,v}$ the line defined by $p_u$ and
$p_v$.
\item{\textbf{I-P2}:} Let the face-vertex $f$ be drawn at point $p_f$. \emph{The perpendicular from point $p_f$ to line $\ell_{u,v}$
intersects the line segment $p_up_v$}. Let $p$ be the point of
intersection.
\item{\textbf{I-P3}:} \emph{There exists two parallel semi-lines $\ell_u$ and $\ell_v$
passing from $p_u$ and $p_v$, respectively, that define a semi-strip
to the right of segment $p_up_v$ that does not intersect the drawing
constructed so far.} Denote this empty semi-strip by $R_{u,v}$.
\end{description}

We proceed to describe how to recursively produce a drawing for tree
$T_g$ and its corresponding subgraph of $G$ so that the overall
drawing is a $\GDual$ drawing for $G$ and its dual. Refer to
Figure~\ref{fig:outerplanar-genaral-case}. Let $p_g$ be a point in
semi-strip $R_{u,v}$ that also belongs to the perpendicular line to
line-segment $p_up_v$ that passes from point $p$. Thus, the segment
corresponding to edge $(f,g)$ of the dual crosses at right-angle the
segment corresponding to edge $(u,v)$ of $G$, as required. If $g$ is
a leaf, i.e., all the edges of face $f$ except $(u,v)$ are edges of
the external face, then we can easily draw the remaining edges of
face $g$ as a polyline of the appropriate number of points that goes
around $p_g$ and connects $p_u$ and $p_v$.

\begin{figure}[h!tb]
  \centering
  \begin{minipage}[b]{.68\textwidth}
     \centering
     \subfloat[\label{fig:outerplanar-genaral-case}{}]
     {\includegraphics[width=\textwidth]{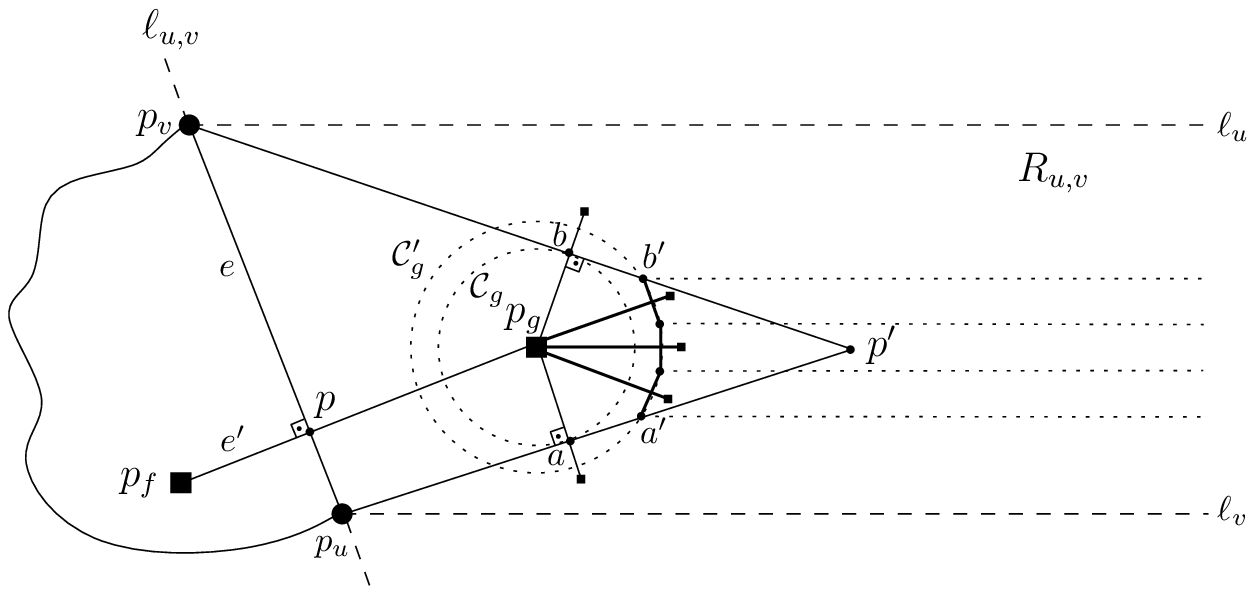}}
  \end{minipage}
  \hfill
  \begin{minipage}[b]{.31\textwidth}
     \centering
     \subfloat[\label{fig:outerplanar-beginning}{}]
     {\includegraphics[width=\textwidth]{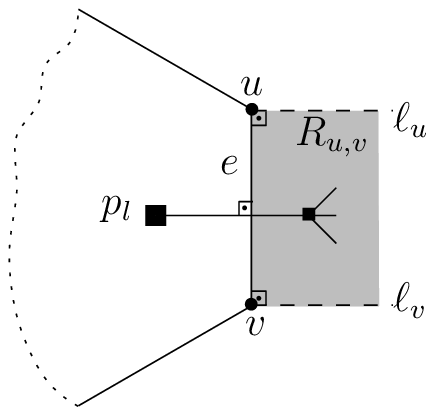}}
  \end{minipage}
  \caption{(a)~The recursion step of our algorithm, (b)~The initial step of our algorithm.}
  \label{fig:ata-graph}
\end{figure}

Consider now the more interesting case where $g$ is not a leaf in
the dual tree of $G$. In this case, we draw two circles, say
$\mathcal{C}_g$ and $\mathcal{C}^{\prime}_g$, centered at $p_g$ that
both lie entirely within semi-strip $R_{u,v}$ and do not touch
neither line $\ell_u$ nor line $\ell_v$. Assume that circle
$\mathcal{C}^{\prime}_g$ is the external of the two circles. From
point $p_u$ draw the tangent to circle $\mathcal{C}_g$ and let $a$
be the point it touches $\mathcal{C}_g$ and $a^{\prime}$ be the
point to the right of $a$ where the tangent intersects circle
$\mathcal{C}^{\prime}_g$ (see
Figure~\ref{fig:outerplanar-genaral-case}). Similarly, we define
points $b$ and $b^{\prime}$ based on the tangent from point $p_v$ to
circle $\mathcal{C}_g$.

Let $k \geq 4$ be the number of vertices defining face $g$. The case
where $k=3$ will be examined later. Draw $k-4$ points on the
$(a^{\prime},b^{\prime})$ arc, which is furthest from segment
$p_up_v$. These points, say $\{p_i~|~1\leq i \leq k-4\}$, together
with points $p_u$, $p_v$, $a^\prime$ and $b^\prime$ form face $g$.
Observe that from point $p_g$, we can draw perpendicular lines
towards each edge of the face. Indeed, line segments $p_ga$ and
$p_gb$ are perpendicular to $p_ua^\prime$ and $p_vb^\prime$,
respectively. In addition, the remaining edges of the face are
chords of circle $\mathcal{C}^\prime_g$ and thus, we can always draw
perpendicular lines to their midpoints from the center $p_g$ of the
circle. Now, from each of the newly inserted points of face $g$ draw
a semi-line that is parallel to semi-line $\ell_u$ and lies entirely
in the semi-strip $R_{u,v}$. We observe all invariant properties
stated above hold for each child of face $g$ in the subtree $T_g$ of
the dual of $G$. Thus, our algorithm can be applied recursively.

The case where the number $k$ of vertices defining face $g$ is equal
to $3$ can be easily treated. We simply use the intersection of the
two tangents, say $p'$, as the third point of the triangular face.
We have to be careful so that $p'$ lies inside the semi-strip.
However, we can always select a point $p_g$ close to segment
$p_up_v$ and an appropriately small radius for circle
$\mathcal{C}_g$, so that $p'$ is inside $R_{u,v}$.

Now that we have described the recursive step of the algorithm, it
is easy to define how the recursion begins (see
Figure~\ref{fig:outerplanar-beginning}). We  start from any face of
$G$ that is a leaf at its dual tree, say face $l$. We draw the face
as regular polygon, with face-vertex $l$ mapped to the center, say
$p_l$, of the polygon. Let $e=(u,v)$ be the only edge of the face
that is internal to the outerplane embedding of $G$. Without loss of
generality, assume that $e$ is drawn vertically. Then, draw the
horizontal semi-lines $\ell_u$ and $\ell_v$ from the endpoints of
$e$ in order to define the semi-strip $R_{u,v}$. From this point on,
the algorithm can recursively draw the remaining faces of $G$ and
its dual. \qed
\end{proof}

We note that the produced $\GDual$ drawing of $G$ and its dual $G^*$
simply proves that producing such drawings is feasible. The drawing
is not particularly appealing since the height of the strips quickly
becomes very small. However, it is a starting point towards
algorithms that produce better layouts. Also note that the algorithm
performs a linear number of ``point computations'' since for each
face-vertex of the dual tree the performed computations are
proportional to the degree of the face-vertex. However, the
coordinates of some points may be non-rational numbers.

\section{Conclusion - Open Problems}
\label{sec:conclusion}

In this paper, we introduced and examined geometric simultaneous RAC
drawings. Our study raises several open problems. Among them are the
following:
\begin{enumerate}
  \item What other non-trivial classes of graphs, besides a matching and
  either a path or a cycle, admit a $\GSimRAC$ drawing?
  \item We considered only geometric simultaneous RAC drawings. For
  the classes where $\GSimRAC$ drawings are not possible, study
  drawings with bends.
  \item We demonstrated by an example that if two graphs always admit a geometric
  simultaneous drawing, it is not necessary that they also admit a
  $\GSimRAC$ drawing. Finding a class of graphs (instead of a particular
  graph) with this property would strengthen this result.
  \item Obtain more appealing $\GDual$ drawings for an outerplanar
  graph and its dual. Study the required drawing area.
\end{enumerate}




\begin{thebibliography}{10}
\providecommand{\url}[1]{\texttt{#1}}
\providecommand{\urlprefix}{URL }

\bibitem{ACBDFKS09}
Angelini, P., Cittadini, L., {Di Battista}, G., Didimo, W., Frati,
F.,
  Kaufmann, M., Symvonis, A.: On the perspectives opened by right angle
  crossing drawings. In: Proc. of 17th International Symposium on Graph
  Drawing. LNCS, vol. 5849, pp. 21--32 (2009)

\bibitem{AGKN10}
Angelini, P., Geyer, M., Kaufmann, M., Neuwirth, D.: On a tree and a
path with
  no geometric simultaneous embedding. In: Proc. of 18th International
  Symposium on Graph Drawing. pp. 38--49. LNCS (2010)

\bibitem{ABS11}
Argyriou, E.N., Bekos, M.A., Symvonis, A.: The straight-line
{R}{A}{C} drawing
  problem is {N}{P}-hard. In: Proc. of 37th International Conference on Current
  Trends in Theory and Practice of Computer Science, (SOFSEM). pp. 74--85. LNCS
  (2011)

\bibitem{AFKMT10}
Arikushi, K., Fulek, R., Keszegh, B., Moric, F., Toth, C.: Graphs
that admit
  right angle crossing drawings. In: Graph Theoretic Concepts in Computer
  Science. LNCS, vol. 6410, pp. 135--146 (2010)

\bibitem{BCDE07}
Brass, P., Cenek, E., Duncan, C.A., Efrat, A., Erten, C.,
Ismailescu, D.,
  Kobourov, S.G., Lubiw, A., Mitchell, J.S.B.: On simultaneous planar graph
  embeddings. Computational Geometry: Theory and Applications  36(2),  117--130
  (2007)

\bibitem{BS93}
Brightwell, G., Scheinerman, E.R.: Representations of planar graphs.
SIAM
  Journal Discrete Mathematics  6(2),  214--229 (1993)

\bibitem{CvKLMSV11}
Cabello, S., v.Kreveld, M.J., Liotta, G., Meijer, H., Speckmann, B.,
Verbeek,
  K.: Geometric simultaneous embeddings of a graph and a matching. JGAA  15(1),
   79--96 (2011)

\bibitem{DGDLM10}
{Di Giacomo}, E., Didimo, W., Liotta, G., Meijer, H.: Area, curve
complexity,
  and crossing resolution of non-planar graph drawings. In: Proc. of 17th
  International Symposium on Graph Drawing. LNCS, vol. 5849, pp. 15--20 (2009)

\bibitem{DEL09}
Didimo, W., Eades, P., Liotta, G.: Drawing graphs with right angle
crossings.
  In: Proc. of 12th International Symposium, Algorithms and Data Structures.
  LNCS, vol. 5664, pp. 206--217 (2009)

\bibitem{DEL10}
Didimo, W., Eades, P., Liotta, G.: A characterization of complete
bipartite
  graphs. Information Processing Letters  110(16),  687--691 (2010)

\bibitem{EL11}
Eades, P., Liotta, G.: Right angle crossing graphs and 1-planarity.
In: 27th
  European Workshop on Computational Geometry (2011)

\bibitem{EK05}
Erten, C., Kobourov, S.G.: Simultaneous embedding of planar graphs
with few
  bends. JGAA  9(3),  347--364 (2005)

\bibitem{EGJPSS07}
Estrella-Balderrama, A., Gassner, E., J{\"u}nger, M., Percan, M.,
Schaefer, M.,
  Schulz, M.: Simultaneous geometric graph embeddings. In: Proc of 15th
  International Symposium on Graph Drawing. pp. 280--290. LNCS (2007)

\bibitem{FKK09}
Frati, F., Kaufmann, M., Kobourov, S.G.: Constrained simultaneous
and
  near-simultaneous embeddings. JGAA  13(3),  447--465 (2009)

\bibitem{GKV09}
Geyer, M., Kaufmann, M., Vrto, I.: Two trees which are
self-intersecting when
  drawn simultaneously. Discrete Mathematics  309(7),  1909--1916 (2009)

\bibitem{Hu07}
Huang, W.: Using eye tracking to investigate graph layout effects.
In: Proc. of
  IEEE Pacific Visualization Symposium. pp. 97--100. IEEE (2007)

\bibitem{HHE08}
Huang, W., Hong, S.H., Eades, P.: Effects of crossing angles. In:
Proc. of IEEE
  Pacific Visualization Symposium. pp. 41--46. IEEE (2008)

\bibitem{vK10}
van Kreveld, M.: The quality ratio of {RAC} drawings and planar
drawings of
  planar graphs. In: Proc. of 18th International Symposium on Graph Drawing.
  pp. 371--376. LNCS (2010)

\end{thebibliography}



\end{document}